\documentclass[11pt,a4paper]{article} 
\usepackage{jheppub}

\usepackage{epsfig,multicol,bbm}
\usepackage{graphicx}
\usepackage{amsmath}
\usepackage{bm}
\usepackage{multirow}
\allowdisplaybreaks[1]

\newcommand{\nl}{\nonumber\\}
\newcommand{\ol}{\overline}
\newcommand{\bea}{\begin{eqnarray}}
\newcommand{\eea}{\end{eqnarray}}

\newcommand\fverbdo{\egroup\medskip\noindent%
			\fbox{\unhbox\fverbbox}\ }
\newcommand\fverbit{\egroup\item[\fbox{\unhbox\fverbbox}]}
\newbox\fverbbox

\newcommand{\be}{\beta}
\newcommand{\ga}{\gamma}
\newcommand{\la}{\lambda}
\newcommand{\si}{\sigma}
\renewcommand{\th}{\theta}

\newcommand{\De}{\Delta}

\def\gtsim{\mathrel{\hbox{\raise0.2ex
\hbox{$>$}\kern-0.75em\raise-0.9ex\hbox{$\sim$}}}}
\def\ltsim{\mathrel{\hbox{\raise0.2ex
\hbox{$<$}\kern-0.75em\raise-0.9ex\hbox{$\sim$}}}}

\preprint{KIAS-O14003}

\title{3.5 keV X-ray Line Signal from Dark Matter Decay in Local $U(1)_{B-L}$ Extension of Zee-Babu Model}

\author[a]{Seungwon Baek}

\affiliation[a]{School of Physics and Open KIAS Center, KIAS,\\ 85 Hoegiro Dongdaemun-gu, Seoul 130-722, Korea}

\emailAdd{swbaek@kias.re.kr}

\abstract{We consider a local $U(1)_{B-L}$ extension of Zee-Babu model to explain the recently
observed 3.5 keV X-ray line signal. The model has three Standard model (SM)-singlet Dirac fermions with different $U(1)_{B-L}$ charges.
A complex scalar field charged under $U(1)_{B-L}$ is introduced to break the $U(1)_{B-L}$ symmetry.
After $U(1)_{B-L}$ symmetry breaking a remnant discrete symmetry stabilizes the lightest state of the Dirac fermions, 
which can be a stable dark matter (DM). 
The second lightest state, if mass splitting with the stable DM is about 3.5 keV, decays dominantly to the stable DM
and 3.5 keV photon through two-loop diagrams, explaining the X-ray line signal. 
Two-loop suppression of the decay amplitude makes its  lifetime much 
longer than the age of the universe and it can be a decaying DM candidate in large parameter region.
We also introduce a real scalar field which is singlet under both the SM and $U(1)_{B-L}$ and can explain the current 
relic abundance of the Dirac fermionic DMs. If the mixing with the SM Higgs boson is small, 
it does not contribute to DM direct detection. The main contribution to the scattering of DM off atomic nuclei comes from 
the exchange of $U(1)_{B-L}$ gauge boson, $Z'$, and is suppressed below current experimental bound when $Z'$ mass is
heavy  ($\gtrsim 10$ TeV). If the singlet scalar mass is about 0.1--10 MeV, DM self-interaction can be large enough to solve small 
scale structure problems in simulations with the cold DM, such as, the core-vs-cusp problem and too-big-to-fail problem.
}
\keywords{  }
\begin{document} 
\maketitle
\section{Introduction\label{sec:intro}}
Although the existence of dark matter (DM) is now well-established from various observations,
the nature of DM(s) is still not well-known.
Thus the search for DM interactions, especially non-gravitational ones, is one of the hot topics in theoretical and
experimental physics.
The recently observed 3.5 keV X-ray line signal in a stacked spectrum of galaxies
and clusters of galaxies~\cite{Xray_exp}, if it is confirmed\footnote{There is some dispute over the
existence of the signal~\cite{dispute}.}, can be a strong  hint for the non-gravitational 
DM interaction. The conventional scenario for the X-ray line in terms of DM models 
is the decay of sterile neutrino with mass $m_s=7.06 \pm 0.5$ keV into a 3.5 keV photon and an active neutrinos.
The observed flux~\cite{Xray_exp,Frandsen:2014lfa} 
\bea
\Phi_{\rm X-ray} \propto n_s \Gamma_s &=& 1.39 \times 10^{-22}\, {\rm s}^{-1} \sin^2 2\th \left(m_s \over {\rm keV}\right)^5 
\frac{\rho_{\rm DM}}{m_s} \nl
&=& (1.5 \times 10^{-25} - 2.7 \times10^{-24})\, {\rm cm^{-3} s^{-1}},
\label{eq:flux}
\eea
can be explained by mixing angle given by $\sin^2 2\th =(2-20) \times 10^{-11}$.
There are already many other scenarios considered on the nature of DMs which can be the source of the
X-ray line~\cite{recent}.

It would be interesting to consider a model with an interplay between DM and other sectors of the SM, for example, 
the neutrino sector~\cite{Ma:2006km,Lindner:2011it,Baek:2012ub}. 
Then measurement of one sector may predict or constrain the other sector.
One of these scenarios has been studied in~\cite{Baek:2012ub}.
In \cite{Baek:2012ub}, we introduced scalar dark matter coupled to the Zee-Babu model
which generates neutrino masses radiatively at two-loop level~\cite{Zee-Babu}. 
We showed that the model can successfully explain Fermi-LAT 130 GeV gamma-ray line.

In this paper we gauge the global $U(1)_{B-L}$ symmetry of \cite{Lindner:2011it,Baek:2012ub}.
To cancel the gauge anomaly we need to introduce three right-handed neutrinos $N_{R_i} (i=1,2,3)$ with $B-L=-1$.
We also introduce a complex scalar field $\varphi$ with $B-L=2$ which breaks the local $U(1)_{B-L}$ symmetry
when $\varphi$ gets vacuum expectation value (VEV), $v_\varphi$.
The $\varphi$ field also generates the soft lepton number breaking term of original Zee-Babu model
dynamically~\cite{Lindner:2011it,Baek:2012ub}. 
The $U(1)_{B-L}$ symmetry would allow the Yukawa interaction $\ell H N_{R_i}$ and Majorana mass terms 
$ N_{R_i} N_{R_i} \varphi$, which would generate neutrino masses  through the usual seesaw mechanism after $U(1)_{B-L}$ symmetry is broken.
Since we want to generate the neutrino masses only through Zee-Babu mechanism~\cite{Zee-Babu},
we forbid the above Yukawa interaction by introducing a global $Z_2$ symmetry
under which only $N_{R_i}$ are odd
and all other particles are even.

As a consequence, the three right-handed neutrinos would not decay and could be potential dark matter candidates,
if the $Z_2$ were unbroken symmetry.
Since all the three right-handed neutrinos have the same $B-L$ charges, however, 
the $U(1)_{B-L}$-gauge interactions are {\it flavor-diagonal} in the mass eigenstate basis,
{\it i.e.} there is no flavor-changing $U(1)_{B-L}$ gauge interactions. And we could not generate processes of 
the form $N_{R_j} \to N_{R_i} \gamma (i \not= j)$ for the X-ray line, even if $N_R$'s were dark matter candidates.
In fact, the global symmetry can be broken by quantum gravity effect~\cite{Kallosh:1995}, which will generate Planck
mass suppressed higher order operators mediating rapid decay of right-handed neutrinos, thus eliminating them from
the list of dark matter candidates.

We introduce Dirac fermionic dark matter candidates $\psi_i (i=1,2,3)$  to explain the X-ray line signal.
The $\psi_i$ are neutral under the SM gauge group
but charged under the local $U(1)_{B-L}$ symmetry. They are vector-like under $U(1)_{B-L}$ symmetry
and gauge anomaly is not generated.
If we assign the $U(1)_{B-L}$ charges of $\psi_i$ fields in such a way that
$\Delta Q_{\psi} \equiv Q_{\psi_2} - Q_{\psi_1} = Q_{\psi_3} -Q_{\psi_2}=2$, {\it off-diagonal} Yukawa interactions, $\ol{\psi_1} \psi_2 \varphi^*$
and $\ol{\psi_2} \psi_3 \varphi^*$, are allowed. After $\varphi$ gets VEV,
off-diagonal terms in the mass matrix of $\psi$'s are generated, which induces the dark-flavor-changing $Z'$ couplings 
at tree level. And flavor-changing radiative decay of DM is allowed.
We can also see that a discrete symmetry remains after $U(1)_{B-L}$ symmetry is broken. 
This {\it local} discrete symmetry guarantees absolute stability of the lightest state of $\psi_i$~\cite{local_DM},
as opposed to the global symmetry which can be broken by quantum gravity.

Finally we introduce a real scalar field $\eta$ which is singlet under both the SM gauge group and $U(1)_{B-L}$.
It has even parity under $Z_2$.
It can couple to the Dirac dark matter fields as $-\eta (y_1 \ol{\psi_1} \psi_1 +y_2 \ol{\psi_2} \psi_2 +y_3 \ol{\psi_3} \psi_3 )$,
while it does not couple to the right-handed neutrinos at tree-level.
The current relic abundance of dark matters mainly come from 
$\psi_i\overline{\psi}_i \to \eta \eta$ process.
{
For the coupling $y_i \sim 0.1$ and $m_\psi \sim 1 \,{\rm TeV}$, we obtain the
annihilation cross section
\bea
\langle \sigma v (\psi_i\overline{\psi}_i \to \eta \eta)\rangle \sim { y_i^2 \over m_\psi^2} \sim 3 \times 10^{-26} \,
{\rm cm^3/sec},
\eea
which gives the correct relic abundance in our universe.
}

If $\eta$ has a mass in the range $m_\eta \sim 0.1-10$ MeV, the elastic scattering cross section of DM can be large enough,
$\sigma/m_\psi \sim 0.1-10 \,{\rm cm^2/g}$ at $v\sim 10\, {\rm km/s}$ relevant for dwarf galaxy scale~\cite{SIDM},
to solve small scale structure problems of cold DM,
such as the core-vs-cusp problem and too-big-to-fail problem.
The contribution of $\eta$ to direct detection cross section of DM can be suppressed below the current experimental
bound because the mixing with the SM Higgs is constrained to be small ($\sim 10^{-5}$) due to non-observation of invisible Higgs decay.

We show that transition magnetic dipole operator (TMDO) $\ol{\psi'_1} \si_{\mu\nu} \psi'_2 F^{\mu\nu}/\Lambda$ 
($\psi'_i$ are mass eigenstates)
can be generated by two-loop diagrams involving Zee-Babu scalars, $\varphi$ scalar, and $B-L$ gauge boson.
The heavier state $\psi'_2$  can decay into the lighter state and photon through this TMDO.
If the mass difference between the two states is about $3.5$ keV, we can explain the observed
X-ray line signal. Since the TMDO is generated at two-loop level, the effective cut-off scale $\Lambda$ of the operator can be
very high, even if all the particles running inside the loop have (sub-)electroweak scale masses. 
As a consequence $\psi'_2$ can live much longer than the age of the universe and can be a decaying DM candidate.

{
In our model there appear some small parameters, such as
$v_\eta/v_\varphi$, $\Delta m_{21}/m_\psi$, {\it etc.}, 
which seems to be fine-tuning at first sight. However, we will show that they
are technically natural in the sense of 't Hooft~\cite{tHooft}: 
\bea
\text {\it ``A parameter is naturally small if setting it to
  zero increases the symmetry of the theory.''}.
\label{itm:tHooft}
\eea
}

The rest of the paper is organized as follows. In Section~\ref{sec:model}, we introduce our model and
show analytic formula for the Wilson coefficient of  TMDO.
In Section~\ref{sec:decayingDM}, we present decaying DM scenario and numerical results.
We conclude in Section~\ref{sec:Conclusions}.

\section{The model}
\label{sec:model}
The model contains two electrically charged Zee-Babu scalar fields 
$h^+$, $k^{++}$, a SM-singlet complex  
dark scalar $\varphi$, a singlet real scalar $\eta$, three right-handed neutrinos $N_{R_i} (i=1,2,3)$
and  three SM-singlet Dirac fermion dark matter candidates $\psi_i$ in addition to
the SM fields. 
In Table~\ref{tab:B-L}, we show the charge assignments of the fields under $U(1)_{B-L}$, and $Z'_2$. 
\begin{table}[htb]
\begin{center}
\begin{tabular}{|c|c|c|c|c|c|c|c|}
\hline
Fields &    $q_{i}$ & $\ell_i$ &$h^+, k^{++}$ & $\varphi$ & $\eta$ & $N_{R_i}$ & $\psi_i$ \\
\hline\hline
$B-L$ &    $1/3$ & $-1$                 &  $2$         & $2$  & $0$&   $-1$ & $1/3,7/3,13/3$ \\
\hline
$Z_2$ &    $+$ & $+$                 &  $+$                                & $+$  & $+$ &  $-$ & $\pm$ \\
\hline
\end{tabular}
\end{center}
\caption{The assignment of $B-L$ charges ($i=1,2,3$).}
\label{tab:B-L}
\end{table}
The Lagrangian for the model can be written as~\cite{Zee-Babu}
\bea
{\cal L}&=& {\cal L}_{\rm SM}+{\cal L}_{\rm Zee-Babu} +{\cal L}_{\rm kin}+ {\cal L}_\Psi -V, \nl
{\cal L}_{\rm Zee-Babu} &=& f_{ab} l_{aL}^{Ti} C l_{bL}^{j}  \epsilon_{ij} h^+
+ h^\prime_{ab} l_{aR}^{T} C l_{bR}^{j}  k^{++} + {h.c}, \nl
{\cal L}_{N_R} &=&  \ol{N_{R_i}}  i \ga^\mu D_\mu N_{R_i}  -{1 \over 2} \Big(\la_{N_{ij}} \varphi \ol{N^c_{R_i}} N_{R_j} + h.c.\Big)\nl
{\cal L}_\Psi &=& \ol{\psi_i}  i \ga^\mu D_\mu \psi_i- m_{\psi_i} \ol{\psi_i} \psi_i 
-f_{12} \Big(\ol{\psi_1} \psi_2 \varphi^* + \ol{\psi_2} \psi_1 \varphi \Big)
-f_{23} \Big(\ol{\psi_2} \psi_3 \varphi^* + \ol{\psi_3} \psi_2 \varphi \Big) \nl
&& -\eta( y_1 \ol{\psi_1} \psi_1 + y_2 \ol{\psi_2} \psi_2 + y_3  \ol{\psi_3} \psi_3), \nl
{\cal L}_{\rm kin} &=& |{\cal D}_\mu h^+|^2 +|{\cal D}_\mu k^{++}|^2 +|{\cal D}_\mu \varphi|^2
{ +{1 \over 2} \left(\partial_\mu \eta\right)^2
+\sum_{i=1}^{3} \ol{\psi_i} i \gamma^\mu {\cal D}_\mu \psi_i}
 -{1 \over 4 } \hat{Z}^\prime_{\mu\nu} \hat{Z}^{\prime \mu\nu}
-{\sin\chi \over 2 } \hat{Z}'_{\mu\nu} \hat{B}^{\mu\nu},\nl
V&=& \mu_H^2 H^\dagger H +\mu_\varphi^2 \varphi^* \varphi + {1 \over 2} \mu_\eta^2 \eta^2+\mu_h^2 h^+ h^- + \mu_k^2 k^{++} k^{--} \nl
&& + (\lambda_\mu \varphi k^{++} h^- h^-  + h.c) \nl
&& + \lambda_H (H^\dagger H)^2  + \lambda_\varphi (\varphi^*
\varphi)^2 + {1 \over 4}\lambda_\eta \eta^4 + \lambda_h (h^+ h^-)^2  
 + \lambda_k (k^{++} k^{--})^2   \nl
&& +\lambda_{H\varphi} H^\dagger H \varphi^* \varphi
+\lambda_{Hh} H^\dagger H h^+ h^- +\lambda_{Hk} H^\dagger H k^{++}k^{--}  
 \nl
&& +\lambda_{\varphi h} \varphi^* \varphi h^+ h^-
+\lambda_{\varphi k}  \varphi^* \varphi k^{++} k^{--}
+\lambda_{hk} h^+ h^- k^{++}k^{--}   \nl
&& +{1 \over 2} \lambda_{H \eta} H^\dagger H \eta^2+{1 \over 2} \lambda_{\varphi \eta} \varphi^* \varphi \eta^2 + \cdots,
\label{eq:Lag}
\eea
where ${\cal D}_\mu =\partial_\mu + i \hat{e} Q \hat{A}_\mu + i \hat{g}_{Z'} Q' \hat{Z}'_\mu$,
and $\hat{B}_{\mu\nu}$, $\hat{Z}'_{\mu\nu}$ are the field strength tensors of $U(1)_Y$, $U(1)'_{B-L}$ gauge field,
respectively.  
The ellipses in the scalar potential $V$ denote the interaction terms involving $\eta$,
which are not important in our analysis and we neglect those terms. 
The above Lagrangian allows Majorana neutrino masses at two-loop level~\cite{Zee-Babu}.
Here the Lepton number (or the $B-L$ number) is broken spontaneously by the $\lambda_\mu$ term,
thus this term provides dynamical origin of soft lepton number breaking term in the original
Zee-Babu model.
We refer the reader to \cite{Zee-Babu} for more details and to \cite{recent_ZB} for the recent analysis of Zee-Babu model.
We just note that we do not have additional constraints from the flavor changing neutral processes
in the quark sector or the charged lepton flavor violations because the $Z'$ boson coupling to the SM fermions
are flavor universal and also we take $Z'$ mass to be very heavy ($M_{Z'} \gtrsim 10$ TeV).
The interplay between the neutrino masses and the dark matter phenomenology comes
from the Zee-Babu scalars, $h^+, k^{++}$, which enter both the neutrino masses and dark matter decays
through loop diagrams. Both prefer the electroweak scale masses of Zee-Babu scalars and can be a target
of LHC searches for new particles beyond the SM.

Note that neither $\ell H N_{R_i}$ term nor $\ell H \psi_i$ term is not allowed due to $Z_2$ parity and $B-L$ charge assignment, respectively. 
Consequently the usual seesaw mechanism does not contribute to the generation of the neutrino masses, 
making the Zee-Babu mechanism dominant one.

The charge assignment allows dark-flavor-changing off-diagonal interactions, $\ol{\psi_1} \psi_2 \varphi^*$ and 
$\ol{\psi_2} \psi_3 \varphi^*$. After $U(1)_{B-L}$ symmetry breaking, these terms induce off-diagonal terms in the
$\psi$ mass matrix, making the rotation from the interaction basis to mass basis non-trivial.
Since the $B-L$ charges of $\psi_i$ are not universal, dark-flavor-changing $Z'$- and $\phi$-interactions and TMDO
are generated.
In addition, due to these terms,  after $\varphi$ gets VEV, the $U(1)_{B-L}$ symmetry is not completely broken, but there remains
discrete $Z_6$ symmetry under which $h^+, k^{++}, \varphi$, and $\eta$ are invariant and
\bea
 q (\psi) &\rightarrow& e^{i 2 \pi/6} q (\psi), \quad l (N_R) \rightarrow e^{i 3 \pi/6} l (N_R), 
\eea
where all the $\psi_i$ have the same $Z_6$ charges\footnote{We set $\Delta Q_\psi =2$ for this purpose.}.
The color-$SU(3)$ symmetry of the SM and $Z_6$ symmetry forbids $\psi$ from decaying\footnote{Mathematically,
the equation $2+3 n=0 \,({\rm mod} \,6)$ does not have any solution for integer $n$.}.
Since the origin of this $Z_6$ symmetry is local gauge symmetry, it is not broken by quantum gravity, and it guarantees
the absolute stability of the dark matters $\psi'_1$ which is the lightest mass eigenstate of $\psi_i$.
The $Z_2$ parity of $\psi_i$ can be either $+$ or $-$ as indicated in Table~\ref{tab:B-L}, and the parity
does not make much phenomenological differences.
If $\psi$'s have odd parity, for example, dimension-6 operator $\psi_1 \psi_1 \psi_1 N_R$ is allowed and it helps $N_R$ decay.

{ The $\hat{Z}^\prime$ boson can mix with the SM hypercharge boson through loop-diagrams involving particles
charged under both $U(1)$'s, generating
the kinetic mixing term $-\sin\chi \hat{Z}'_{\mu\nu} \hat{B}^{\mu\nu}/2$ in ({\ref{eq:Lag}})~\cite{Holdom,Babu:1997st}.
The current bound on the mixing angle $\chi$ is at most ${\cal O}(10^{-3})$, depending on the mass of $Z'$~\cite{Essig:2013}. 
However, we note that this mixing term cannot generate the TMDO for $\psi'_2 \to \psi'_1 \gamma$. The argument
goes as follows.  For simplicity we consider only the photon and $\hat{Z}^\prime$ boson mixing,
$-\sin\chi \hat{Z}'_{\mu\nu} \hat{F}^{\mu\nu}/2$
($F_{\mu\nu}=\partial_\mu A_\nu -\partial_\nu A_\mu$ is the photon field strength tensor), 
although it is straightforward to generalize to the case of ({\ref{eq:Lag}}).
 The kinetic energy terms including the above mixing term are transformed into canonical
form by a non-unitary transformation,
\bea
\left(
\begin{array}{c}
\hat{A} \\
\hat{Z}'
\end{array}
\right)
=\left(
\begin{array}{cc}
1 & -\tan\chi \\
0 & \sec\chi
\end{array}
\right)
\left(
\begin{array}{c}
A \\
Z'
\end{array}
\right).
\label{eq:non-unitary}
\eea
The resulting kinetic energy and mass terms of photon and $Z'$ are written as
\bea
\Delta {\cal L} = -{1 \over 4} F_{\mu\nu} F^{\mu\nu} -{1 \over 4} Z'_{\mu\nu} Z^{'\mu\nu}
+ {1 \over 2} M_{Z'}^2 Z'_{\mu} Z^{'\mu},
\label{eq:gauge}
\eea
where we assumed the scalar field giving mass to $Z'$ carries only the extra $U(1)$ charge as in our model.
We can see that the Lagrangian in (\ref{eq:gauge}) has additional $SO(2)$-symmetry $(A_\mu, Z'_\mu)-$space, when $M_{Z'}=0$.
However, this is not the case for the massive $Z'$ case because it makes the photon massive.
This shows the uniqueness of the transformation (\ref{eq:non-unitary}) when $Z'$ is massive. In this case
the covariant derivative is given by
\bea
D_\mu &=& \partial_\mu + i \hat{e} Q \hat{A}_\mu + i \hat{g}_{Z^\prime}  Q_{Z'} \hat{Z}^\prime_\mu \nl 
&=& \partial_\mu + i \hat{e} Q A_\mu + {i \over \cos\chi} (\hat{g}_{Z^\prime}  Q_{Z'} - \hat{e} Q \sin\chi) Z^\prime_\mu.
\eea
This proves our argument that the DM carrying only dark charge does not couple to the photon for the
massive $Z'$ case even if there is mixing between $\gamma$ and $Z'$.
From the above equation we can identify $\hat{e}$ with physical electromagnetic charge $e$ in our case~\cite{Babu:1997st}.
This should be contrasted with the massless $Z'$ case where milli-charged dark matter is allowed\footnote{In this
case the $SO(2)$ rotation should be exploited in such a way that the electromagnetic charge $Q$ couples fully to
the photon~\cite{Huh:2007zw}. Then the covariant derivative looks like 
$D_\mu=\partial_\mu + i \left({\hat{e} \over \cos\chi} Q-\hat{g}_{Z'} Q_{Z'} \tan\chi\right) A_\mu + i \hat{g}_{Z'} Q_{Z'} X_\mu$,
which shows that the dark sector particles can couple to the photon field.}
}

{ Since the kinetic mixing term does not generate TMDO and also $\chi \lesssim O(10^{-3})$, its effect on
TMDO is at most subdominant and we set $\chi=0$ for simplicity.}
We identify $A \equiv \hat{B} c_W + \hat{W}^3 s_W$, $Z \equiv -\hat{B} s_W + \hat{W}^3 c_W$ 
and $Z^\prime \equiv \hat{Z}^\prime$ with the photon, $Z$-boson, and $B-L$ gauge boson, respectively.
Their masses are $0$, $M_{Z}=91.1876$ GeV~\cite{Beringer:1900zz}, and $M_{Z'}$, respectively.
The $Z^\prime$ mass is strongly constrained by the LEPII experiment~\cite{Cacciapaglia}
\bea
 \frac{M_{Z^\prime}}{g_{Z'}} > 7 \, {\rm TeV},
\label{eq:Zprime_mass_bound}
\eea
at 99 \% confidence level.

The neutral scalar fields $h$, $\phi$ and $\eta$ can also mix with each other after they obtain vacuum expectation values:
\bea
H = \frac{1}{\sqrt{2}} \left(\begin{array}{c} 0 \\ v_H + h\end{array} \right), \quad
\varphi = \frac{1}{\sqrt{2}}(v_\varphi + \phi), \quad
\eta = v_\eta + n.
\eea
The mixing of $h$ with the SM-singlet scalars $\phi$ and $n$ is strongly constrained by the
invisible Higgs decay width~\cite{Baek:2011aa}, although it can help Higgs potential
stable at high energy~\cite{vac_sta}. And we (almost) neglect their mixings\footnote{We will allow, however,
small mixing between $H$ and $\eta$, when  we consider the decay of $n$.}.
In other words $h$, $\phi$ and $n$ are considered as mass eigenstates with masses $m_h, m_\phi$ and $m_n$.
From (\ref{eq:Zprime_mass_bound}), we expect $v_\varphi \gtrsim {\cal O}(10\, {\rm TeV})$ for $g_{Z'} \lesssim {\cal O}(1)$.
We set $m_n \sim 0.1-10$ MeV, from which we expect 
$v_\eta \sim {\cal  O}(10-100\, {\rm MeV})$.
{  The large hierarchy between $v_\varphi$ and $v_\eta$ can be ascribed
to the hierarchy between $\mu_\varphi^2$ and $\mu_\eta^2$, if we
assume the mixing parameter $\lambda_{\varphi \eta}$ is much smaller
than the quartic couplings, $\lambda_\varphi, \lambda_\eta$. The small
$\mu_\eta^2$ compared with $\mu_\varphi^2$, however, is still natural
according to (\ref{itm:tHooft}). It is because the Lagrangian
(\ref{eq:Lag}) has additional symmetry, {\it i.e.} scale invariance
for the transformation
\bea
 x \to x e^\sigma, \quad \eta(x) \to e^{-\sigma} \eta(x e^{-\sigma}),
\eea
in the limit $\mu_\eta^2 \to 0$.}


We should get non-trivial mixing in $\psi_i$ states to generate TMDO.
After $U(1)_{B-L}$ symmetry breaking,  the mass terms of the Dirac dark fermions are given by\footnote{
There is also small contribution from $\eta \overline{\psi_i} \psi_i$ interactions in the diagonal part.
But since $v_\eta \sim 10-100$ MeV, they are small and we absorb them to $m_{\psi_i}$}
\bea
{\cal L}_{\text{$\psi$ mass}} &=& -
\left(\begin{array}{ccc}\ol{\psi_1} & \ol{\psi_2}  & \ol{\psi_3} \end{array}\right)
\left(\begin{array}{ccc}
m_{\psi_1} & \frac{f_{12} v_\varphi}{\sqrt{2}} & 0\\
\frac{f_{12} v_\varphi}{\sqrt{2}}  & m_{\psi_2} & \frac{f_{23} v_\varphi}{\sqrt{2}} \\
0 &\frac{f_{23} v_\varphi}{\sqrt{2}}   & m_{\psi_3}  
\end{array}
\right)
\left(\begin{array}{c}
{\psi_1} \\
{\psi_2} \\
{\psi_3}
\end{array}
\right).
\eea
%
Assuming $f$'s are real, the mass eigenstates $\psi_i^\prime$ are obtained by an orthogonal rotation
\bea
\left(
\begin{array}{c}
\psi_1 \\ \psi_2 \\ \psi_3
\end{array}
\right)
=
O
\left(
\begin{array}{c}
\psi'_1 \\ \psi'_2 \\ \psi'_3
\end{array}
\right),
\eea
with the corresponding masses $m_{\psi_i^\prime}$.
The lightest $\psi_1^\prime$ is absolutely stable due to the local $Z_6$ symmetry and become a DM candidate.  
{ We take $m_{\psi'_1} \sim {\cal O}(1) \, {\rm TeV}$, because this gives not only the correct relic density, but
the necessary self-scattering cross section to solve the small scale structure problems of the CDM,
when the coupling of DM with the light scalar is of order one.}
The $\psi'_2$ can decay into $\psi_1^\prime$ and a photon through the TMDO, which can explain the 3.5 keV X-ray line signal.
It can also be a DM component if its lifetime is much longer than the age of the universe.
To get 3.5 keV X-ray line in the decay process $\psi_2^\prime \to \psi_1^\prime \gamma$ we fix the mass difference
\bea
 \Delta m_{21} \equiv m_{\psi_2^\prime} - m_{\psi_1^\prime} 
= \frac{2 m_{\psi_2^\prime} E_\gamma}{m_{\psi_2^\prime} + m_{\psi_1^\prime}} \simeq E_\gamma =3.5\, {\rm keV},
\eea
where we assumed $m_{\psi_i^\prime} \gg 3.5\, {\rm keV}$. 
From $\De m_{21} \approx\sqrt{(m_{\psi_1} - m_{\psi_2})^2 + 2 f_{12}^2 v_\varphi^2}$, we see
$\sqrt{2} |f_{12}| v_\varphi \le 3.5$ keV, and 
\bea
|f_{12}| \lesssim 10^{-10}
\eea
 for $v_\varphi \sim {\cal O}(\rm 10 \, TeV)$. 
Thus the Yukawa coupling constant $f_{12}$ is very small. 
{ It should be noted that, if we set $\Delta m_{21}=0$, we get additional $U(2)$ symmetry in the 
$(\psi_1, \psi_2)-$flavor space. This means that
his small $\Delta m_{21}/m_{\psi_i}$ parameter is technically natural
according to the 't Hooft's naturalness criterion (\ref{itm:tHooft}).}
Naturally we also expect $(2,3)$-component of the mass matrix
is much smaller than the diagonal components. We assume the following hierarchy for the 
parameters: $ m_{\psi'_i} (\sim{\cal O}(1 \, {\rm TeV})) \gg f_{23} v_\varphi \gg f_{12} v_\varphi (\sim {\cal O}(1 \, {\rm keV}))$.
Now we can readily diagonalize the $\psi$ mass matrix. We do this in two steps: first, we diagonalize the $2 \times 2$
submatrix  for $(\psi_1, \psi_2)$ exactly without perturbation and then we use the 1st order perturbation to diagonalize the full matrix.
In the end we get\footnote{We use the convention $O = O_{23} O_{13} O_{12} $ with $O_{ij}$ a rotation matrix in $i-j$ plane.}
\bea
O &\simeq& \left(\begin{array}{ccc}
c_{12} & s_{12} & s_{13} \\
-s_{12} & c_{12} & s_{23} \\
s_{12} s_{23} & -c_{12} s_{23} & 1 
\end{array}\right), \nl
\tan2 \theta_{12}&=&\frac{\sqrt{2} f_{12} v_\varphi}{m_{\psi_2}-m_{\psi_1}}, \quad
s_{13} \simeq \frac{f_{23} v_\varphi s_{12}/\sqrt{2}}{m_{\psi'_3}-m_{\psi'_1}}, \quad
s_{23} \simeq \frac{f_{23} v_\varphi c_{12}/\sqrt{2}}{m_{\psi'_3}-m_{\psi'_2}}, \nl
m_{\psi^\prime_{1,2}} &=&{1 \over 2} \left(m_{\psi_1}+m_{\psi_2}\mp \sqrt{(m_{\psi_1}-m_{\psi_2})^2+2 f_{12}^2 v_\varphi^2}\right),
\quad m_{\psi'_3} \simeq m_{\psi_3},
\eea
where $c_{ij} (s_{ij})$'s are abbreviation of $\cos \theta_{ij}  (\sin \theta_{ij}) $ and we have assumed $\theta_{13}, \theta_{23} \ll 1$.
The parameters $m_{\psi_i}$, $f_{12}$ and $f_{23}$ in the Lagrangian can be expressed in terms of the mass eigenvalues and
mixing angles for which we take as inputs as follows:
\bea
  m_{\psi_i} &=& \sum_{k=1}^3 O_{ik}^2 m_{\psi'_k} \nl
  f_{ij (i<j)} &=& \frac{\sqrt{2}}{v_\varphi} \sum_{k=1}^3 O_{ik} O_{jk} m_{\psi'_k}.
\eea
However all the mixing angles are not independent because $f_{13}=0$, which gives the constraint
\bea
 s_{13}=-\frac{t_{12} c_{23} s_{23} \Delta m_{32}}{\Delta m_{31}-s_{23}^2 \Delta m_{32}} \simeq -t_{12} c_{23} s_{23},
\eea
where $t_{12} =\tan\theta_{12}$, $\Delta m_{ij} =m_{\psi'_i}-m_{\psi'_j}$ and the approximation holds for $s_{23} \ll 1$ and 
$\Delta m_{31} \simeq \Delta m_{32}$.


\begin{figure}
\center
\includegraphics[width=5cm]{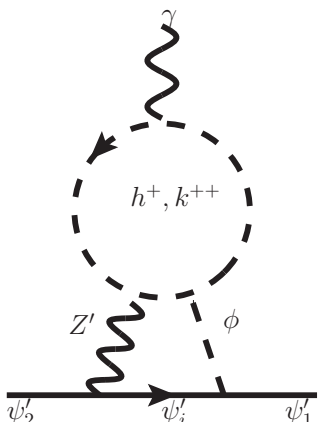}
\caption{A Barr-Zee type two-loop diagram for $\psi'_2 \to \psi'_1 + \gamma$.}
\label{fig:Barr-Zee}
\end{figure}
The effective operator for magnetic transition $\psi'_2 \to \psi'_1 \gamma$, is given by
\bea
{\cal L}_{\rm eff} &=& {1 \over \Lambda} \overline{\psi'_1} \sigma_{\mu\nu} \psi'_2 F^{\mu\nu},
\label{eq:MDO}
\eea
It is generated by so-called ``Barr-Zee'' type two-loop diagrams~\cite{Arhrib:2001xx} 
with topology shown in Fig.~\ref{fig:Barr-Zee}. The state $\psi'_2$ decays almost
100\% via (\ref{eq:MDO})~\cite{Frandsen:2014lfa}.
Given that $\chi$~\cite{Essig:2013lka} and $h-\phi(n)$~\cite{Baek:2011aa} mixing are strongly constrained and the Barr-Zee type diagrams
are generated even in the limits where those mixings vanish, we can consider the effects of non-vanishing mixings
as small perturbations. 
The leading contribution of two-loop Barr-Zee type diagrams to $1/\Lambda$ is obtained to
be
\bea
{1 \over \Lambda}&\simeq&-\sum_{s=h^+,k^{++}}\frac{8 e g^2_{Z'}  \Delta Q_\psi   Q_s Q_{s}^\prime  \lambda_{\varphi s}
\delta^2 \cos 2\theta_{12} s_{13} s_{23} }{(4\pi)^4}  \nl
&\times&\int_0^1 dx \int [d\be]
\frac{x \be_4^2 m_{\psi'_3}^3}
{\left(\be_1 m_{Z'}^2 + \be_2 m_\phi^2 +\be_3 m_s^2/(x(1-x)) +\be_4^2 m_{\psi'_1}^2\right)^2},
\label{eq:d_M}
\eea
where $[d\be] \equiv d\be_1 d\be_2 d\be_3 d\be_4 \delta(1-\be_1 -\be_2 -\be_3 -\be_4)$, 
$\Delta Q_\psi = Q_{\psi'_3}-Q_{\psi'_2}= Q_{\psi'_2}-Q_{\psi'_1}=2$, $\delta=\Delta m_{31}/m_{\psi'_3}$ 
and we neglected small contribution proportional to $\Delta m_{21} (\simeq 3.5 {\rm keV})$.
{The scale $\Lambda$ is roughly given by
\bea
\Lambda &\sim& 10^8 \times \left(0.2 \over \delta \right)^2 \times \frac{\max(m_{Z'}^4, m_\phi^4, m_{h^+}^4, m_{k^{++}}^4,m_{\psi'_1}^4)}{m_{\psi'_3}^3},
\label{eq:Lambda}
\eea
for $g_{Z'} \sim 0.3$, $\theta_{12} \sim \theta_{23} \sim 0.2$.
For $m_{Z'} \sim 10$ TeV, $m_{\psi'_3} \sim 1$ TeV, $\delta \sim 0.2$, we get 
$\Lambda \sim 10^{15}$ GeV, which can explain the 3.5 keV X-ray line signal (See (\ref{eq:Lambda_exp})).
We note that large $\Lambda$ ($\sim 10^{15}$ GeV) is generated mainly from two-loop suppression factor,
although the typical dimensionless parameters are of order $0.1-1$ and the involved particle masses are of order $1 - 10$ TeV.
}

\section{Decaying dark matter scenario}
\label{sec:decayingDM}

In principle the observed X-ray line can be explained in two scenarios in our model. In the first scenario,
the dark matters scatter inelastically into excited states, $\psi'_1 + \psi'_1 (\overline{\psi'_1}) \to \psi'_2 + \psi'_2 (\overline{\psi'_2})$, 
$ \psi'_1 + \psi'_2 (\overline{\psi'_2})$, 
followed by
the decay of excited states, $\psi'_2 (\overline{\psi'_2})\to \psi'_1 (\overline{\psi'_1})+ \gamma$, via TDMO.
In the second scenario, the lifetime of $\psi'_2$ is much longer than the age of the universe and
it can be a decaying dark matter candidate. When it decays, it can also give X-ray line signal via
$\psi'_2 (\overline{\psi'_2})\to \psi'_1 (\overline{\psi'_1})+ \gamma$.

In  the exciting dark matter scenario the decay rate 
\bea
\Gamma_\psi\equiv \Gamma_{\psi'_2\to \psi'_1 \gamma} =\frac{4  (\De m)^3}{\pi \Lambda^2}
\eea
should be larger than the upward scattering rate, 
$\Gamma_{\psi\psi}\equiv n_{\psi'_1} \sigma_{\psi'_1  \psi'_1  \to \psi'_2  \psi'_2 } v_{\rm rel}$~\cite{Frandsen:2014lfa}.
Since we need
\bea
(\sigma_{\psi'_1  \psi'_1  \to \psi'_2  \psi'_2 } v_{\rm rel}) \times {\rm BR}(\psi'_2\to \psi'_1 \gamma) \approx
(1.7 \times 10^{-22} - 3.0 \times 10^{-21}) {\rm cm^3 s^{-1}} \left(m_{\psi'_1} \over {\rm GeV} \right)^2,
\eea
to explain the X-ray signal~\cite{Frandsen:2014lfa}, the condition, $\Gamma_\psi > \Gamma_{\psi\psi}$, corresponds to
$\Lambda  \lesssim$ $10^{15} \,{\rm GeV} \,({\rm GeV}/m_{\psi'_1})^{1/2}$ or 
$\tau_{\psi'_2}  \lesssim  10^{22} \, {\rm s} \,({\rm GeV}/m_{\psi'_1})$.
We have checked that we need rather large $g_{Z'}(\sim 5)$ and/or $\lambda_{\varphi h(k)} (\sim 5)$ and
perturbativity assumed in obtaining (\ref{eq:d_M}) is not well-justified. This is understandable because our TMDO
is generated at two-loop level and loop-suppression factor is very large.
And we do not consider the possibility of exciting dark matter scenario further.

In the decaying DM scenario, the lifetime of $\psi'_2$ should be longer than the age of the universe,
which gives the constraint
\bea
\Lambda > 6.12 \times 10^{12} \, {\rm GeV}.
\label{eq:longevity}
\eea
after $\psi'_{1,2}$ are decoupled from the thermal plasma at temperature $T_f \approx m_{\psi'_{1,2}}/20$,
the ratio of the $\psi'_2$ to the $\psi'_1$ number density is almost fixed to be
\bea
\frac{n_{\psi'_2}}{n_{\psi'_1}} \approx e^{-\De m/T_f} \approx 1, 
\eea
for $\De m(=3.5\, {\rm keV}) \ll T_f$.
To explain the X-ray line signal we require 
\bea
 n_{\psi'_2} \Gamma_{\psi'_2 \to \psi'_1 \gamma} = {1 \over 2} n_{\rm DM} \Gamma_{\psi'_2 \to \psi'_1 \gamma}
\eea
should in the range given in (\ref{eq:flux}). This corresponds to
\bea
\Lambda = (6.94 \times 10^{14} - 2.95 \times 10^{15}) \left(m_{\psi'_2} \over {\rm GeV}\right)^{-1/2} \, {\rm GeV}.
\label{eq:Lambda_exp}
\eea

\begin{figure}
\center
\includegraphics[width=7cm]{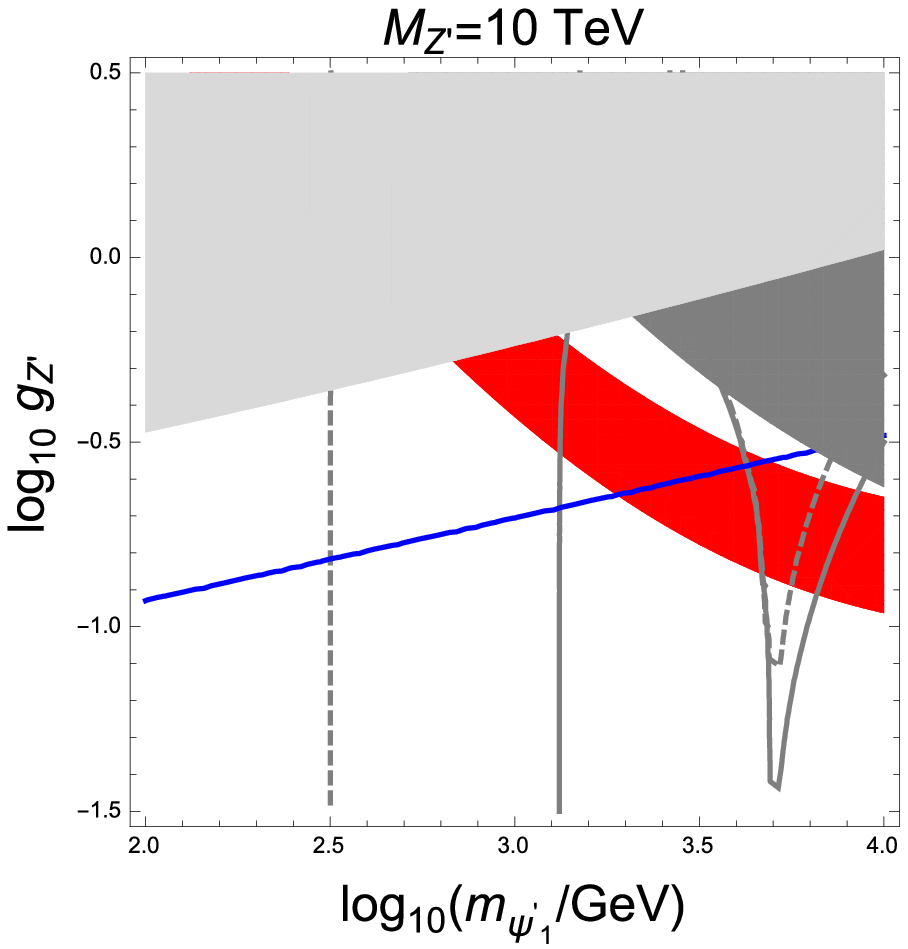}
\includegraphics[width=7cm]{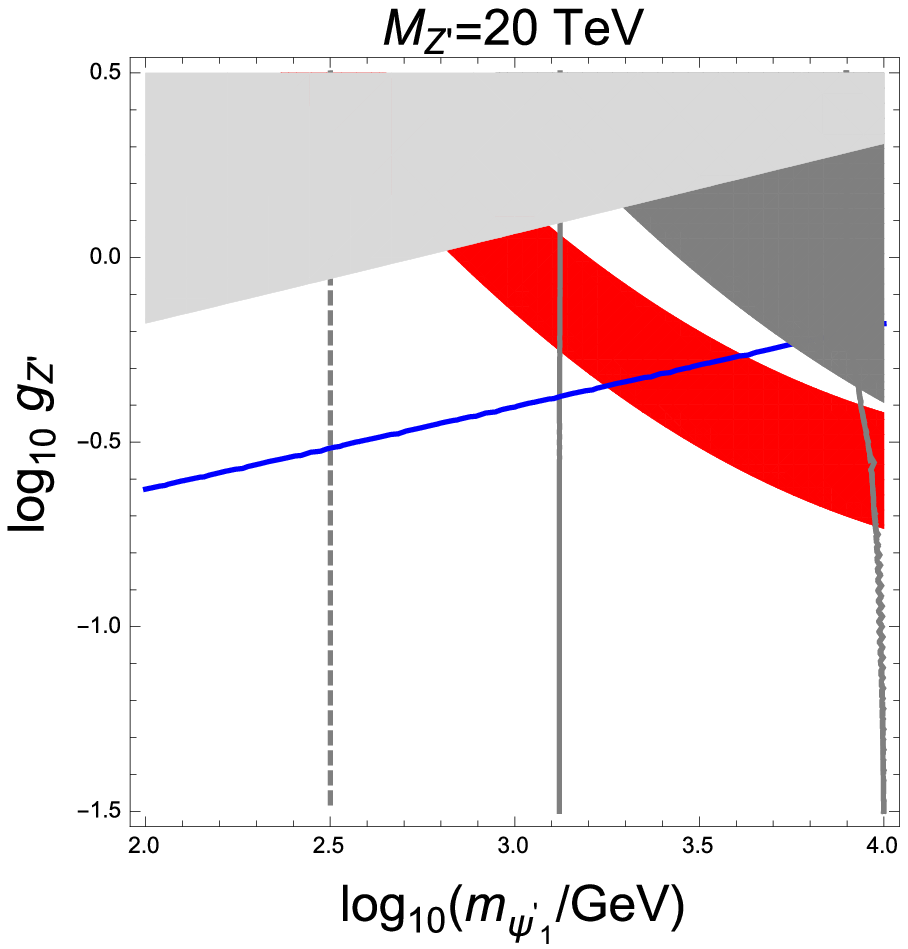}
\caption{Plots in $(m_{\psi'_1},g_{Z'})$-plane. The red-colored region can explain the 3.5 keV X-ray line signal. The
dark gray region is excluded because the lifetime of $\psi'_2$ is shorter than the age of the universe.
The light gray region is excluded by LUX DM direct detection experiment. The blue line is the sensitivity the next
XENONO1T experiment can reach. The black solid (dashed) line gives the correct relic abundance of DM for $y_i=2(1)$. 
For the left (right) plot we set $M_{Z'}=10 (20)$ TeV.}
\label{fig:X-ray}
\end{figure}

In Fig.~\ref{fig:X-ray}, the red-colored region satisfies this and explains the observed
X-ray line signal in the $(m_{\psi'_1},g_{Z'})$-plane. For the left (right) panel we have taken $M_{Z'}=10 \,(20)$ TeV.
For other parameters we have fixed\footnote{Although we take small $\theta_{12}$, it can be
${\cal O}(1)$ in general. The consequent change on $\Lambda$ can be easily seen from (\ref{eq:d_M}). }, $\delta=0.2$, $\th_{12}=\th_{23} =0.2$, 
$m_\phi=  m_{h^+}= m_{k^{++}}=1$ TeV, $\lambda_{\varphi h} =\lambda_{\varphi k} =1$.
We have checked that the signal region is not very sensitive to the mass parameters,
$m_\phi, m_{h^+}$ and $m_{k^{++}}$.
The black solid (dashed) lines satisfy the observed relic abundance of dark matters,
$\Omega_\psi h^2 =0.1199 \pm 0.0027$~\cite{Planck:2013}, for $y_i=2 \, (1)$.
The vertical lines come from the annihilation channel $\psi_{1(2)} \overline{\psi_{1(2)}} \to n n$ and therefore
are sensitive to the Yukawa couplings $y_i$. There are also resonance regions when $m_{\psi'_1} \approx M_{Z'}/2$.
The region with dark gray color is excluded because it does not satisfy the longevity of the decaying DM, 
(\ref{eq:longevity}).
The light grey region is excluded by LUX DM direct search experiment~\cite{LUX} and blue line show
the sensitivity of future DM experiment XENON1T~\cite{XENON1T}.
In our case the direct detection of DM is dominated by $Z'$ boson exchange diagram even though $Z'$ is very heavy,
$M_{Z'}=10 (20) \, {\rm TeV}$.
{We note that $m_{h^+}= m_{k^{++}}=1$ TeV can easily evade the constraints from the lepton flavor violating
processes with $f_{ab} , h'_{ab} \sim {\cal O}(0.01)$, while 
still being able to explain neutrino masses (for example, see Herrero-Garcia, {\it et.al.} in~\cite{recent_ZB}).
}
Although there is no direct signature for our scenario, we need relatively light, electroweak scale, Zee-Babu scalars
$h^+$ and $k^{++}$, which can be searched for at LHC 14 TeV.

{
Sizable invisible Higgs decay width can also support the existence of $\eta$
because, as we will see, $n$\footnote{More precisely, the mass eigenstate with $n$ component the largest. It can
be understood from the context whether $n$ represents interaction eigenstate or mass eigenstate.} 
does not decay inside particle detectors and appears as
invisible signal due to its long lifetime of ${\cal O}(1)$ sec.
The LHC Higgs measurements constrains the branching fraction of the invisible Higgs decay~\cite{invisible_Higgs},
\bea
{\cal B}_h^{\rm inv} < 0.58 ~~~@ \,95 \% \, {\rm CL},
\eea
which implies the invisible Higgs decay width $\Gamma^{\rm inv} < 1.38 \Gamma^{\rm SM} \approx 5.52 \,{\rm MeV}$.
The decay width of  invisible Higgs decay mode,  $h \to n n$, is given by
\bea
 \Gamma(h \to n n) =\frac{\lambda_{H\eta}^2 v^2}{32 \pi m_h}.
\eea
The current bound on the decay width then gives bound on $\lambda_{H\eta}$,
\bea
\lambda_{H\eta} < 0.034.
\eea
For light $n$, the mixing angle between $n$ and $h$ is  constrained to be
\bea
\alpha_{H\eta} \approx \frac{\lambda_{H\eta} v v_\eta}{m_h^2} 
< 5.4 \times 10^{-5} \left(v_\eta \over 100 {\rm MeV}\right).
\eea
Thus we can see that there is still much room for invisible Higgs decay into light scalar $n$,
although the mixing angle between the Higgs and $n$ is strongly constrained.
}

{ The singlet $n$ can also decay into the SM particles very fast, thus not causing any cosmological problems. For example, when
$m_n >2 m_e$, the $n$ can decay into an electron-positron pair through mixing with the Higgs field. The decay width is
given by
\bea
\Gamma(n \to e^+e^-) = \frac{G_F \sin^2 \alpha_{\eta H}}{4 \sqrt{2} \pi} m_n m_e^2 \beta_e^3,
\eea
where $\alpha_{\eta H}$ is the mixing angle between $\eta$ and $H$, and $\beta_e=(1-4 m_e^2/m_n^2)^{1/2}$.
For $m_n=10\, {\rm MeV}$, $\alpha_{\eta H}=10^{-4}$, the lifetime of $n$ becomes about
 $0.04 \, {\rm sec}$. Since $n$ can decay long before ~1 sec, it does not affect big bang neucleosysthesis (BBN).
As a subdominant decay channel, we also have $n \to \gamma \gamma$. We note that this channel does not require the mixing 
of $n$ with $H$. The $n$ can decay into two photons through the loop processes where $h^+$ and $k^{++}$ particles
are running inside the loop. The decay width for this two photon channel is given by~\cite{Baek:2012ub},
\bea
\Gamma(n \to \gamma\gamma) =\frac{\alpha_{\rm em}^2}{64 \pi^3 m_n}
\Bigg|
(\mu_{\eta h} +2 \lambda_{\eta h} v_\eta) (1-\tau_h f(\tau_h))
+4 (\mu_{\eta k} +2 \lambda_{\eta k} v_\eta) (1-\tau_k f(\tau_k))
\Bigg|^2,
\eea
where $\tau_i =4 m_i^2/m_n^2\, (i=h^+, k^{++})$. The loop function is given as
\bea
f(\tau)=
\left\{
\begin{array}{ll}
-\frac{1}{4} \left(\log\frac{1+\sqrt{1-\tau}}{1-\sqrt{1-\tau}}-i \pi\right)^2 & (\tau <1) \\
\arcsin^2 \frac{1}{\sqrt{\tau}} & (\tau \geq 1).
\end{array}
\right.
\eea
For $v_\eta =50\, {\rm MeV}, m_n=10\, {\rm MeV}, \mu_{\eta h}=\mu_{\eta k}=10^{-4} \, {\rm MeV}, \lambda_{\eta h}=\lambda_{\eta k}=10^{-2},
m_{h^+}=m_{k^{++}}=1\, {\rm TeV}$,  we get the partial decay width $\Gamma(n \to \gamma\gamma) \simeq 6 \times 10^{-33} \, {\rm GeV}$,
corresponding to the lifetime $\simeq 1.2 \times 10^8 \, {\rm sec}$.
}

\begin{figure}
\center
\includegraphics[width=8cm]{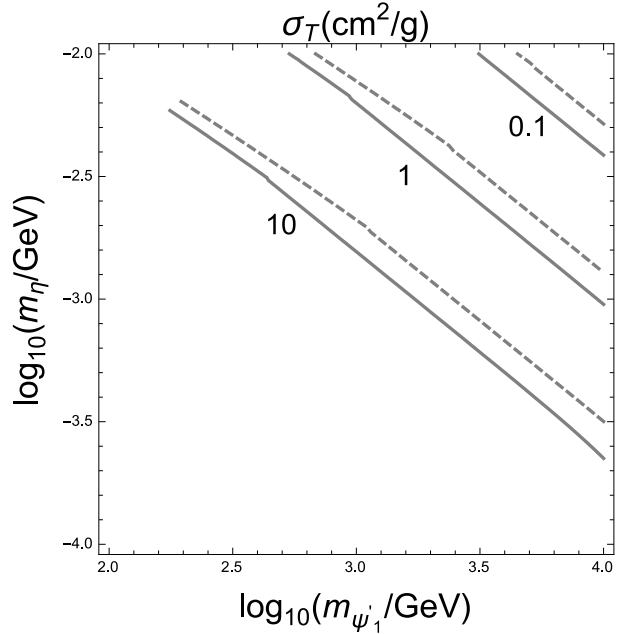}
\caption{Contour plot of $\sigma_T=0.1, 1, 10 (\rm cm^2/g)$ (from right above to left below) 
which may solve the core-vs-cusp problem and too-big-to-fail problem. The solid (dashed) lines correspond
to $\alpha_y=1/4\pi (2^2/4\pi)$.}
\label{fig:sigma_T}
\end{figure}

For light $\eta$ particle ($m_\eta \sim 1-10 \, {\rm MeV}$), $\eta$-exchanging (in)elastic self-interacting processes
$\psi'_{1(2)},\psi'_{1(2)} \to \psi'_{1(2)},\psi'_{1(2)}$ can be strong. When they have cross sections
\bea
 \sigma_T/m_{\psi'_1} \sim 0.1 - 10 \, {\rm cm^2/g},
\eea
we can solve the small scale structure problems such as core-vs-cusp problem and too-big-to-fail problem in our model.
In our model, we need relatively large ($y_i \sim {\cal O}(1)$) Yukawa coupling of $\eta$ with the DM, to get the correct
relic density\footnote{We can significantly reduce the necessary $y_i$ if we include the Sommerfeld enhancement
effect~\cite{Sommerfeld}.}. In this case the self-interaction typically occurs in the non-perturbative 
($\alpha_y m_{\psi'} /m_\eta \gtrsim 1$ with $\alpha_y \equiv y^2/4\pi$) classical regime
($m_{\psi'} v_{\rm rel}/m_\eta \gg 1$), where the transverse cross section $\sigma_T$ is given by~\cite{SIDM}
\bea
\sigma_T =\int d\Omega (1-\cos\theta) \frac{d\sigma}{d\Omega}
=\left\{\begin{array}{ll}
{4 \pi \over m_\eta^2} \be^2 \ln(1+\be^{-1}) & \text{for} \;\be \lesssim 10^{-1} \\
{8 \pi \over m_\eta^2} \be^2 /(1+1.5 \be^{1.65}) & \text{for} \;10^{-1} \lesssim \be \lesssim 10^{3} \\
{\pi \over m_\eta^2} \left(\ln\be +1 -{1 \over 2} \ln^{-1} \be\right)^2 & \text{for} \; \be \gtrsim 10^{3},
\end{array}
\right.
\eea
where $\be \equiv 2 \alpha_y m_\eta/(m_\psi v_{\rm rel}^2)$. In Fig.~\ref{fig:sigma_T}, we show $\sigma_T$ contour
plots in $(m_{\psi'_1},m_{\eta})$-plane for $\alpha_y=1/4\pi$ (solid line) and  $\alpha_y=2^2/4\pi$ (dashed line).
We have taken $v_{\rm rel}=10 \, {\rm km/s}$ which is relevant for dwarf galaxies.
We can see that the DM scattering cross section can be in the $0.1-10 \, {\rm cm^2/g}$  range for $m_{\psi'_1}=0.1 - 10$ TeV
and $m_\eta=0.1-10$ MeV.

{As mentioned in the Introduction, the discrete symmetry $Z_2$ can be broken by quantum gravity effect, which will
result in rapid decay of right-handed neutrinos $N_{R_i}$. Now let us address the decay of $N_{R_i}$ in more detail.
The breaking of global $Z_2$ symmetry by gravity would generate Planck mass suppressed higher dimensional
operators~\cite{Hamaguchi:2008ta}
\bea
\frac{1}{M_{\rm Pl}} \ell_i H N_{R_j} \eta, 
\quad \frac{1}{M_{\rm Pl}^2} N_{R_i} \ell_j \ell_k \bar{e}_l, 
 \quad \frac{1}{M_{\rm Pl}^2} N_{R_i} \bar{d}_j \bar{d}_k \bar{u}_l, 
\eea
where $i,j,k,l=1,2,3$ and $M_{\rm Pl}$ is Planck mass. 
Assuming order one coupling, the dimension five operator mediates the decay of $N_{R_i}$ dominantly through
three-body decay, giving their lifetime
\bea
\tau_R \sim \left(1 \, {\rm PeV} \over m_{R_i}\right)^3 0.1 \,{\rm sec}.
\eea
If the right-handed neutrinos are PeV scale, they can decay before BBN occurs ($\sim 1$ sec), causing no cosmological problems.
And only $\psi_i$'s remain as the dominant component of DM in the current universe.
}

\section{Conclusions\label{sec:Conclusions}}
We extended the Zee-Babu model for neutrino masses to have $U(1)_{B-L}$  gauge symmetry and 
to incorporate Dirac dark matters to explain the X-ray line signal. 
We also introduced $U(1)_{B-L}$ breaking scalar, singlet scalar,
and right-handed neutrinos.
The charges of the particle content are assigned 
in such a way that after the $U(1)_{B-L}$ breaking scalar getting VEV
the local $U(1)_{B-L}$ symmetry is broken down to a discrete symmetry. The lightest Dirac dark fermion $\psi'_1$
whose mass is TeV scale
transforms non-trivially under this discrete symmetry and becomes stable.

{
The heavier $\psi'_2$ particle can decay almost 100\% through the magnetic dipole transition operator
$\overline{\psi'_1} \si_{\mu\nu} \psi'_2 F^{\mu\nu}/\Lambda$.
Since this operator is generated at two-loop so-called Barr-Zee diagrams, the cut-off scale
$\Lambda$ is very high $\sim 10^{15}$ GeV and the lifetime of $\psi'_2$ is much longer than the age of the universe.
And $\psi'_2$ can be a decaying dark matter candidate.
If $\De m_{21} =m_{\psi'_2}-m_{\psi'_1} \simeq 3.5$ keV,
the recently claimed X-ray line signal~\cite{Xray_exp} can be accommodated for wide range of dark matter masses.
}

The relic abundance of dark matters in the current universe can also be explained 
by the dark matter annihilation into two singlet scalars and also by the $Z'$-resonance annihilation.
Although our $Z'$ is very heavy $\gtrsim 10$ TeV, it can still mediate the dark matter scattering off atomic
nuclei at the level that can be probed at the next generation dark matter direct search experiments.
The singlet scalar can be very light  ($m_\eta =0.1-10$ MeV) and mediate strong self-interactions of dark matters
with cross section $\sigma_T = 0.1 -10 \, {\rm cm^2/g}$,
which can solve small scale structure problems, such as the core-vs-cusp problem and the too-big-to-fail problems,
of the standard $\Lambda$CDM model.

{ The small mass difference $\Delta m_{21}$ and the small VEV of $\eta$ are technically natural in the sense
of 't Hooft. The singlet scalar and the right-handed neutrinos decay fast without causing any cosmological problems.}

\begin{acknowledgments}
The author is grateful to Pyungwon Ko, Wan-Il Park, Chaehyun Yu for useful discussions. This work was supported by in part NRF Grant 2012R1A2A1A01006053.
\end{acknowledgments}



\begin{thebibliography}{999}
\bibitem{Xray_exp}
 E.~Bulbul, M.~Markevitch, A.~Foster, R.~K.~Smith, M.~Loewenstein and S.~W.~Randall,
  arXiv:1402.2301 [astro-ph.CO];
%
A.~Boyarsky, O.~Ruchayskiy, D.~Iakubovskyi and J.~Franse,
  arXiv:1402.4119 [astro-ph.CO].

\bibitem{Frandsen:2014lfa} 
  M.~T.~Frandsen, F.~Sannino, I.~M.~Shoemaker and O.~Svendsen,
  JCAP {\bf 1405}, 033 (2014)
  [arXiv:1403.1570 [hep-ph]].

\bibitem{recent}
H.~Ishida, K.~S.~Jeong and F.~Takahashi,
  arXiv:1402.5837 [hep-ph];
D.~P.~Finkbeiner and N.~Weiner,
  arXiv:1402.6671 [hep-ph];
 T.~Higaki, K.~S.~Jeong and F.~Takahashi,
  arXiv:1402.6965 [hep-ph];
  J.~Jaeckel, J.~Redondo and A.~Ringwald,
  Phys.\ Rev.\ D {\bf 89}, 103511 (2014)
  [arXiv:1402.7335 [hep-ph]];
  H.~M.~Lee, S.~C.~Park and W.~-I.~Park,
  arXiv:1403.0865 [astro-ph.CO];
  R.~Krall, M.~Reece and T.~Roxlo,
  arXiv:1403.1240 [hep-ph];
  K.~Kong, J.~-C.~Park and S.~C.~Park,
  arXiv:1403.1536 [hep-ph];
  K.~-Y.~Choi and O.~Seto,
  arXiv:1403.1782 [hep-ph];
  S.~Baek and H.~Okada,
  arXiv:1403.1710 [hep-ph];
  M.~Cicoli, J.~P.~Conlon, M.~C.~D.~Marsh and M.~Rummel,
  arXiv:1403.2370 [hep-ph];
  F.~Bezrukov and D.~Gorbunov,
  arXiv:1403.4638 [hep-ph];
  C.~Kolda and J.~Unwin,
  arXiv:1403.5580 [hep-ph];
  R.~Allahverdi, B.~Dutta and Y.~Gao,
  arXiv:1403.5717 [hep-ph];
  N.~-E.~Bomark and L.~Roszkowski,
  arXiv:1403.6503 [hep-ph];
  S.~P.~Liew,
  arXiv:1403.6621 [hep-ph];
  P.~Ko, Z.~kang, T.~Li and Y.~Liu,
  arXiv:1403.7742 [hep-ph];
  S.~V.~Demidov and D.~S.~Gorbunov,
  arXiv:1404.1339 [hep-ph];
  F.~S.~Queiroz and K.~Sinha,
  arXiv:1404.1400 [hep-ph];
  E.~Dudas, L.~Heurtier and Y.~Mambrini,
  arXiv:1404.1927 [hep-ph];
  K.~S.~Babu and R.~N.~Mohapatra,
  arXiv:1404.2220 [hep-ph];
  K.~P.~Modak,
  arXiv:1404.3676 [hep-ph];
  J.~M.~Cline, Y.~Farzan, Z.~Liu, G.~D.~Moore and W.~Xue,
  arXiv:1404.3729 [hep-ph];
  H.~M.~Lee,
  arXiv:1404.5446 [hep-ph];
  D.~J.~Robinson and Y.~Tsai,
  arXiv:1404.7118 [hep-ph];
  J.~P.~Conlon and F.~V.~Day,
  arXiv:1404.7741 [hep-ph];
  S.~Baek, P.~Ko and W.~I.~Park,
  arXiv:1405.3730 [hep-ph];
  S.~Chakraborty, D.~K.~Ghosh and S.~Roy,
  arXiv:1405.6967 [hep-ph];
  N.~Chen, Z.~Liu and P.~Nath,
  arXiv:1406.0687 [hep-ph].
  C.~Q.~Geng, D.~Huang and L.~H.~Tsai,
  JHEP {\bf 1408}, 086 (2014)
  [arXiv:1406.6481 [hep-ph]].
C.~-W.~Chiang and T.~Yamada,
  arXiv:1407.0460 [hep-ph];
B.~Dutta, I.~Gogoladze, R.~Khalid and Q.~Shafi,
  arXiv:1407.0863 [hep-ph];
H.~Okada and Y.~Orikasa,
  arXiv:1407.2543 [hep-ph].
  Y.~Farzan and A.~R.~Akbarieh,
  arXiv:1408.2950 [hep-ph];
  G.~Faisel, S.~Y.~Ho and J.~Tandean,
  arXiv:1408.5887 [hep-ph]
  A.~Falkowski, Y.~Hochberg and J.~T.~Ruderman,
  arXiv:1409.2872 [hep-ph];
  S.~Patra and P.~Pritimita,
  arXiv:1409.3656 [hep-ph].

\bibitem{dispute}
  T.~E.~Jeltema and S.~Profumo,
  arXiv:1408.1699 [astro-ph.HE].
  E.~Bulbul, M.~Markevitch, A.~R.~Foster, R.~K.~Smith, M.~Loewenstein and S.~W.~Randall,
  arXiv:1409.4143 [astro-ph.HE].

\bibitem{Ma:2006km} 
  E.~Ma,
  Phys.\ Rev.\ D {\bf 73}, 077301 (2006)
  [hep-ph/0601225].


\bibitem{Lindner:2011it} 
  M.~Lindner, D.~Schmidt and T.~Schwetz,
  Phys.\ Lett.\ B {\bf 705}, 324 (2011)
  [arXiv:1105.4626 [hep-ph]].

\bibitem{Baek:2012ub} 
  S.~Baek, P.~Ko, H.~Okada and E.~Senaha,
  JHEP {\bf 1409}, 153 (2014)
  [arXiv:1209.1685 [hep-ph]].

\bibitem{Zee-Babu} 
 A.~Zee,
  Nucl.\ Phys.\ B {\bf 264}, 99 (1986);  
%
  K.~S.~Babu,
  Phys.\ Lett.\ B {\bf 203}, 132 (1988); 
%
  K.~S.~Babu and C.~Macesanu,
  Phys.\ Rev.\ D {\bf 67}, 073010 (2003)
  [hep-ph/0212058].

\bibitem{Kallosh:1995} 
  R.~Kallosh, A.~D.~Linde, D.~A.~Linde and L.~Susskind,
  Phys.\ Rev.\ D {\bf 52}, 912 (1995)
  [hep-th/9502069].

\bibitem{local_DM} 
  S.~Baek, P.~Ko and W.~-I.~Park,
  JHEP {\bf 1307}, 013 (2013)
  [arXiv:1303.4280 [hep-ph]];
S.~Baek, P.~Ko and W.~-I.~Park,
  arXiv:1311.1035 [hep-ph];
  S.~Baek, P.~Ko and W.~I.~Park,
  arXiv:1407.6588 [hep-ph].

\bibitem{SIDM} 
  S.~Tulin, H.~B.~Yu and K.~M.~Zurek,
  Phys.\ Rev.\ D {\bf 87}, no. 11, 115007 (2013)
  [arXiv:1302.3898 [hep-ph]].


\bibitem{tHooft}
G. 't Hooft, {\it Naturalness, Chiral Symmetry, and Spontaneous Chiral Symmetry Breaking,}
in the {\it Proceedings of the 1979 Carg\`ese Institute on Recent developments in gauge theories,}
G. 't Hooft et al. eds., Plenum Press, New York, U.S.A. (1980).


\bibitem{recent_ZB}
  D.~Aristizabal Sierra and M.~Hirsch,
  JHEP {\bf 0612}, 052 (2006)
  [hep-ph/0609307];
M.~Nebot, J.~F.~Oliver, D.~Palao and A.~Santamaria,
  Phys.\ Rev.\ D {\bf 77}, 093013 (2008)
  [arXiv:0711.0483 [hep-ph]];
D.~Schmidt, T.~Schwetz and H.~Zhang,
  arXiv:1402.2251 [hep-ph];
J.~Herrero-Garcia, M.~Nebot, N.~Rius and A.~Santamaria,
  arXiv:1402.4491 [hep-ph];
H.~N.~Long and V.~V.~Vien,
  Int.\ J.\ Mod.\ Phys.\ A {\bf 29}, no. 13, 1450072 (2014)
  [arXiv:1405.1622 [hep-ph]].

\bibitem{Hamaguchi:2008ta} 
  K.~Hamaguchi, S.~Shirai and T.~T.~Yanagida,
  Phys.\ Lett.\ B {\bf 673}, 247 (2009)
  [arXiv:0812.2374 [hep-ph]].


\bibitem{Holdom} 
  B.~Holdom,
  Phys.\ Lett.\ B {\bf 166}, 196 (1986);

\bibitem{Babu:1997st}
  K.~S.~Babu, C.~F.~Kolda and J.~March-Russell,
  Phys.\ Rev.\ D {\bf 57}, 6788 (1998)
  [hep-ph/9710441];

\bibitem{Huh:2007zw} 
  J.~H.~Huh, J.~E.~Kim, J.~C.~Park and S.~C.~Park,
  Phys.\ Rev.\ D {\bf 77}, 123503 (2008)
  [arXiv:0711.3528 [astro-ph]].

\bibitem{Essig:2013}
  R.~Essig, J.~A.~Jaros, W.~Wester, P.~H.~Adrian, S.~Andreas, T.~Averett, O.~Baker and B.~Batell {\it et al.},
  arXiv:1311.0029 [hep-ph].


\bibitem{Beringer:1900zz} 
  J.~Beringer {\it et al.}  [Particle Data Group Collaboration],
  Phys.\ Rev.\ D {\bf 86}, 010001 (2012).

\bibitem{Cacciapaglia} 
  G.~Cacciapaglia, C.~Csaki, G.~Marandella and A.~Strumia,
  Phys.\ Rev.\ D {\bf 74}, 033011 (2006)
  [hep-ph/0604111].

\bibitem{Baek:2011aa} 
  S.~Baek, P.~Ko and W.~I.~Park,
  JHEP {\bf 1202}, 047 (2012)
  [arXiv:1112.1847 [hep-ph]];
  S.~Choi, S.~Jung and P.~Ko,
  JHEP {\bf 1310}, 225 (2013)
  [arXiv:1307.3948];
  S.~Baek, P.~Ko and W.~I.~Park,
  Phys.\ Rev.\ D {\bf 90}, 055014 (2014)
  [arXiv:1405.3530 [hep-ph]].

\bibitem{vac_sta}
  O.~Lebedev,
  Eur.\ Phys.\ J.\ C {\bf 72}, 2058 (2012)
  [arXiv:1203.0156 [hep-ph]];
J.~Elias-Miro, J.~R.~Espinosa, G.~F.~Giudice, H.~M.~Lee and A.~Strumia,
  JHEP {\bf 1206}, 031 (2012)
  [arXiv:1203.0237 [hep-ph]];
S.~Baek, P.~Ko, W.~I.~Park and E.~Senaha,
  JHEP {\bf 1211}, 116 (2012)
  [arXiv:1209.4163 [hep-ph]];

\bibitem{Arhrib:2001xx} 
S.~M.~Barr and A.~Zee,
  Phys.\ Rev.\ Lett.\  {\bf 65}, 21 (1990)
  [Erratum-ibid.\  {\bf 65}, 2920 (1990)];
  A.~Arhrib and S.~Baek,
  Phys.\ Rev.\ D {\bf 65}, 075002 (2002)
  [hep-ph/0104225].


\bibitem{Essig:2013lka} 
  R.~Essig, J.~A.~Jaros, W.~Wester, P.~H.~Adrian, S.~Andreas, T.~Averett, O.~Baker and B.~Batell {\it et al.},
  arXiv:1311.0029 [hep-ph].


\bibitem{Planck:2013} 
  P.~A.~R.~Ade {\it et al.}  [Planck Collaboration],
  arXiv:1303.5076 [astro-ph.CO].

\bibitem{LUX} 
  D.~S.~Akerib {\it et al.}  [LUX Collaboration],
  Phys.\ Rev.\ Lett.\  {\bf 112}, 091303 (2014)
  [arXiv:1310.8214 [astro-ph.CO]].

\bibitem{XENON1T} 
  E.~Aprile [XENON1T Collaboration],
  Springer Proc.\ Phys.\  {\bf 148}, 93 (2013)
  [arXiv:1206.6288 [astro-ph.IM]].

\bibitem{invisible_Higgs}
  S.~Chatrchyan {\it et al.}  [CMS Collaboration],
  Eur.\ Phys.\ J.\ C {\bf 74}, no. 8, 2980 (2014)
  [arXiv:1404.1344 [hep-ex]].

\bibitem{Sommerfeld} 
  N.~Arkani-Hamed, D.~P.~Finkbeiner, T.~R.~Slatyer and N.~Weiner,
  Phys.\ Rev.\ D {\bf 79}, 015014 (2009)
  [arXiv:0810.0713 [hep-ph]];
  S.~Baek and P.~Ko,
  JCAP {\bf 0910}, 011 (2009)
  [arXiv:0811.1646 [hep-ph]].


\bibitem{BBN} 
  M.~Kawasaki, K.~Kohri and T.~Moroi,
  Phys.\ Lett.\ B {\bf 625}, 7 (2005)
  [astro-ph/0402490];
K.~Jedamzik,
  Phys.\ Rev.\ D {\bf 74}, 103509 (2006)
  [hep-ph/0604251].

\end{thebibliography}
\end{document}